\documentstyle[12pt,fleqn]{article}
\begin{document}
\baselineskip=8mm
\newcounter{s}[section]
\newcounter{e}[equation]
\renewcommand{\thefootnote}{\fnsymbol{footnote}}

\noindent \textbf{Cerenkov Emission by Neutral Particles in
Gravitoelectro-}

\noindent \textbf{magnetic Fields}
\bigskip

\noindent \textbf{X.Q.Li and S.Q.Liu}\footnote{ Department of
Physics, Nanchang University, Nanchang, P.R.China; to whom
correspondence should be addressed:sqliu@ncu.edu.cn  }

{\_}{\_}{\_}{\_}{\_}{\_}{\_}{\_}{\_}{\_}{\_}{\_}{\_}{\_}{\_}{\_}{\_}{\_}{\_}{\_}{\_}{\_}{\_}{\_}{\_}{\_}{\_}{\_}{\_}{\_}{\_}{\_}{\_}{\_}{\_}{\_}{\_}{\_}{\_}{\_}{\_}{\_}{\_}{\_}{\_}{\_}{\_}{\_}{\_}{\_}{\_}{\_}{\_}{\_}{\_}{\_}{\_}{\_}{\_}{\_}{\_}{\_}{\_}{\_}{\_}{\_}{\_}{\_}{\_}{\_}{\_}{\_}{\_}{\_}{\_}{\_}{\_}{\_}{\_}{\_}{\_}{\_}{\_}{\_}{\_}{\_}{\_}{\_}

It is shown that under the post-Newtonian approximation the
Einstein\\
\indent equations can be reduced to the standard Maxwell-type field
equations \\
\indent in a medium; in such a context the Cerenkov emission by a
neutral\\
\indent particle gives large energy loss while the particle moves
at faster than\\
\indent the phase speed of waves in the medium.

 {\_}{\_}{\_}{\_}{\_}{\_}{\_}{\_}{\_}{\_}{\_}{\_}{\_}{\_}{\_}{\_}{\_}{\_}{\_}{\_}{\_}{\_}{\_}{\_}{\_}{\_}{\_}{\_}{\_}{\_}{\_}{\_}{\_}{\_}{\_}{\_}{\_}{\_}{\_}{\_}{\_}{\_}{\_}{\_}{\_}{\_}{\_}{\_}{\_}{\_}{\_}{\_}{\_}{\_}{\_}{\_}{\_}{\_}{\_}{\_}{\_}{\_}{\_}{\_}{\_}{\_}{\_}{\_}{\_}{\_}{\_}{\_}{\_}{\_}{\_}{\_}{\_}{\_}{\_}{\_}{\_}{\_}{\_}{\_}{\_}{\_}{\_}{\_}

\bigskip

\bigskip
\noindent It has often been noted that magnetism can be understood
as the consequence of electrostatics plus Lorentz invariance.
Similarly, Newtonian gravity together with Lorentz invariance in a
consistent way must include a gravitomagnetic field. This is the
case of gravitoelectromagnetic (GEM) form of the Einstein equations
in a medium [1-5]. In such a framework, the refractive index $n
\equiv \sqrt {\varepsilon ^t} $of a medium is often is more than one
[see (\ref{eq27}), below]; note that the statement that only
accelerated particles radiate is not correct when there are waves
with $n > 1$ [6]. Hence the Cerenkov emission by neutral particles
in constant rectilinear motion is unavoidable.

It is well known that in the slow-motion limit of general
relativity, accurate to post-Newtonian (PN) order, where the field
equations can be reduced as [7]:

\begin{equation}  \label{eq1}
\nabla ^2\phi = 4\pi G\rho ,
\end{equation}

\begin{equation}
\nabla ^2{\rm {\bf A}}=16\pi G\rho {\rm {\bf v}}/c;  \label{eq2}
\end{equation}
moreover the harmonic gauge condition and the Lorentz-type force on a unit
mass reduced to

\begin{equation}  \label{eq3}
\nabla \cdot {\rm {\bf A}} + \frac{4}{c}\frac{\partial }{\partial t}\phi = 0,
\end{equation}

\begin{equation}
\frac{d{\rm {\bf v}}}{dt}\approx -\nabla \phi -\frac{1}{c}\frac{\partial
{\rm {\bf A}}}{\partial t}+\frac{{\rm {\bf v}}}{c}\times \nabla \times {\rm
{\bf A}},  \label{eq4}
\end{equation}%
where $\phi $ is the Newtonian gravitational potential, $A/c=g_{i0}$ is the
mixed metric and $G$ is the Newton's constant.

One defines the GEM fields via

\begin{equation}
\label{eq5} {\rm {\bf E}}_g = - \nabla \phi -
\frac{1}{c}\frac{\partial {\rm {\bf A}} }{\partial
t},{\begin{array}{*{20}c}
 \hfill & {\nabla \times {\rm {\bf A}} = {\rm {\bf B}}_g } \hfill \\
\end{array} }
\end{equation}
in direct analogy with electromagnetism; it follows from these
definitions the Maxwell-type field equations in the medium [8, 9]

\begin{equation}
\nabla \times {\rm {\bf B}}_g=\frac{16\pi }c{\rm {\bf j}}_g+\frac 4c\frac{%
\partial {\rm {\bf E}}_g}{\partial t},\ \quad {\nabla \cdot }{\rm {\bf E}}%
_g=4\pi \rho _g,  \label{eq6}
\end{equation}

\begin{equation}
\nabla \times {\rm {\bf E}}_{g}=-\frac{1}{c}\frac{\partial {\rm {\bf B}}_{g}%
}{\partial t},\quad {\nabla \cdot }{\rm {\bf B}}_{g}=0,\qquad  \label{eq7}
\end{equation}%
where $\rho _{g}=-G\rho $ is \textquotedblleft\ matter
density\textquotedblright\ and ${\rm {\bf j}}_{g}=-G\rho {\rm {\bf
v}}$ is \textquotedblleft\ matter current\textquotedblright . Then
Eq.(\ref{eq4}) becomes

\begin{equation}
\frac{d{\rm {\bf v}}}{dt}\approx {\rm {\bf E}}_{g}+\frac{{\rm {\bf v}}}{c}%
\times {\rm {\bf B}}_{g}.  \label{eq8}
\end{equation}%
And all terms of $O({v^{4}}/c^{4})$ are neglected in the above analysis. Now
one decompose the total matter density and current in equation (\ref{eq6}) :
$\rho _{g}=\tilde{\rho}_{g}+\rho _{g0}{,}{\rm {\bf j}}_{g}={\rm {\bf \tilde{j%
}}}_{g}+{\rm {\bf j}}_{g0}$, in which the $\rho _{g0}$ and ${\rm {\bf j}}%
_{g0}$ are the parts of a external field source in the medium, corresponding
to a local additional mass disturbance, for example, by nonlinear
interactions between the fields and the medium, or some test particles. In
this case it is convenient to definite the vector quantity ${\rm {\bf D}}%
_{g}(t,{\rm {\bf r}})$ through ( in what follows one neglects the marks ` $%
\sim $ ' in $\tilde{\rho}_{g}$and ${\rm {\bf \tilde{j}}}_{g}$ for simplicity
)

\begin{equation}
{\rm {\bf D}}_g={\rm {\bf E}}_g+4\pi \int\limits_{-\infty }^t{d{t}^{\prime }}%
{\rm {\bf j}}_g({t}^{\prime },{\rm {\bf r}});  \label{eq9}
\end{equation}
by use of (\ref{eq9}) and the mass continuity equation,

\[
\frac{\partial \rho _\alpha }{\partial t}+\nabla \cdot \rho _\alpha {\rm
{\bf v}}=0\quad (\alpha =g,g_0)\quad ,
\]
(\ref{eq6}) and (\ref{eq7}) are deduced to

\begin{equation}
\nabla \times {\rm {\bf B}}_g=\frac{16\pi }c{\rm {\bf j}}_{g0}+\frac 4c\frac{%
\partial {\rm {\bf D}}_g}{\partial t},\ \quad {\nabla \cdot }{\rm {\bf D}}_g%
{\rm {\bf =4\pi }}\rho _{g0}\ ,  \label{eq10}
\end{equation}

\begin{equation}
\nabla \times {\bf E}_g=-\frac 1c\frac{\partial {\rm {\bf B}}_g}{\partial t}%
,\quad {\nabla \cdot }{\rm {\bf B}}_g=0{\rm {\bf \ .}}  \label{eq11}
\end{equation}

A dimensional analysis of equations (\ref{eq1}), (\ref{eq5}), (\ref{eq9})
and (\ref{eq11}) suggests that it is insightful to use the cgs units, since
these field variables have the following dimensions:

\[
\left[ {D_g}\right] \sim \left[ {E_g}\right] \sim \left[ {B_g}\right] \sim %
\left[ {\nabla \phi }\right] \sim [v]/s,
\]
thus, one has

\[
\left[ {E_g^2/G}\right] \sim \left[ {B_g^2/G}\right] \sim \left[ {\rho
E_g^2/\rho G}\right] \sim \left[ {\rho v^2}\right] ;
\]
and by substituting
\begin{eqnarray}
{\rm {\bf E}} &=&-{\rm {\bf E}}_g/4\sqrt{G},\quad {\rm {\bf B}}=-{\rm {\bf B}%
}_g/4\sqrt{G},\quad {\rm {\bf D}}=-{\rm {\bf D}}_g/4\sqrt{G},  \label{eq12}
\\
{\rm {\bf j}} &=&-{\rm {\bf j}}_g/\sqrt{G}=\sqrt{G}\rho {\rm {\bf v}},\quad
\widehat{\rho }=-\rho _g/\sqrt{G}=\sqrt{G}\rho ,  \label{eq13}
\end{eqnarray}
the GEM field equations can be expressed in the standard form

\begin{equation}
\nabla \times {\rm {\bf B}}=\frac{4\pi }c{\rm {\bf j}}_0+\frac 1c\frac{%
\partial {\rm {\bf D}}}{\partial t},\quad \nabla \times {\bf E}=-\frac 1c%
\frac{\partial {\rm {\bf B}}}{\partial t}\   \label{eq14}
\end{equation}

\begin{equation}
\nabla \cdot {\rm {\bf D}}=\pi \hat{\rho}_0,\quad \nabla \cdot {\rm {\bf B}}%
=0;  \label{eq15}
\end{equation}
then Eqs.(\ref{eq9}) and (\ref{eq8}) become

\begin{equation}  \label{eq16}
{\rm {\bf D}} = {\rm {\bf E}} + \pi \int\limits_{ - \infty }^t {d{t}^{\prime}%
} {\rm {\bf j}}({t}^{\prime},{\rm {\bf r}}).
\end{equation}

\begin{equation}
\quad \frac{d{\rm {\bf v}}}{dt}\approx -4\sqrt{G}\left[ {{\rm {\bf E}}+\frac{%
{\rm {\bf v}}}{c}\times {\rm {\bf B}}}\right] .  \label{eq17}
\end{equation}%
To consider the responses of the medium on the GEM fields, one must
introduce material relation, which describes the GEM properties of the
medium. In view of (\ref{eq16}), the states of the medium not only depend on
a given time-space point ($t,{\rm {\bf r}})$ but also depend on previous
times and at any point of the medium. Hence, by general reasoning in physics
( independent of a specific model for the medium ) one can state that this
is a no-local linear relation in the limit of linear response ; whose
Fourier representation is

\begin{equation}
j_{i}(\omega {\rm {\bf ,k}})=\sigma _{ij}(\omega {\rm {\bf ,k}})E_{j}(\omega
{\rm {\bf ,k}}).  \label{eq18}
\end{equation}%
In special, for common continuous medium, where the \textquotedblleft\
spatial dispersion \textquotedblright\ (dependency on ${\rm {\bf k}})$ is
not important , the relation (\ref{eq18}) is deduced to ($\sigma
_{ij}\rightarrow \sigma \delta _{ij})$

\[
{\rm {\bf j}}(\omega )=\sigma (\omega ){\rm {\bf E}}(\omega ),
\]%
i.e. an \textquotedblleft\ Ohm's gravitational law \textquotedblright\ [10]
, where $\sigma $ is \textquotedblleft gravitational
conductivity\textquotedblright\ [11] ( for details see Refs [10,12]) . Then
by use of Eqs.(\ref{eq16}) and (\ref{eq18}), one has

\begin{equation}
{\rm {\bf D}}_{i}(\omega ,{\rm {\bf k}})=\varepsilon _{ij}(\omega ,{\rm {\bf %
k}})E_{j}(\omega ,{\rm {\bf k}}),  \label{eq19}
\end{equation}%
here $\varepsilon _{ij}$, called the \textquotedblleft dielectric
tensor\textquotedblright , is determined as

\begin{equation}
\varepsilon _{ij}(\omega {\rm {\bf ,k}})=\delta _{ij}+\frac{\pi i}\omega
\sigma _{ij}(\omega {\rm {\bf ,k}}){\quad }(\omega \neq 0).  \label{eq20}
\end{equation}
In fact , if we consider the kinetic description for a
self-gravitating system, on basis of Vlasor equation of the gravity
[13,14], in which the force is Lorentz-type one [Eq.(\ref{eq17})],
one can expand the distribution function $f$ in powers of the field
$E$; and by Fourier transformation

\begin{equation}
{\rm {\bf A}}({\rm {\bf r}},t)=\int {{\rm {\bf A}}(\omega ,{\rm {\bf k}}%
)e^{-i(\omega t-{\rm {\bf k}}\cdot {\rm {\bf r}})}d\omega d{\rm {\bf k}}%
,\quad {\rm {\bf A}}(\omega ,{\rm {\bf k}})}=\int {{\rm {\bf A}}({\rm {\bf r}%
},t)e^{i(\omega t-{\rm {\bf k}}\cdot {\rm {\bf r}})}\frac{d{\em t}d{\rm {\bf %
r}}}{(2\pi )^4}},  \label{eq21}
\end{equation}
it yields the material relation (\ref{eq18}) for the first order current $%
{\rm {\bf j}}\propto \int {{\rm {\bf v}}f_1({\rm {\bf r}},{\rm {\bf v}},t)}d%
{\rm {\bf p}}$ , in which $\sigma _{ij}(\omega {\rm {\bf ,k}})$ is relative
to the particle distribution in the medium .

By applying the Fourier expansion, the field equation (\ref{eq14}) becomes
\begin{eqnarray*}
i({\rm {\bf k}}\times {\rm {\bf B}})_{i} &=&-\frac{i\omega }{c}\varepsilon
_{ij}E_{j}+\frac{4\pi }{c}j_{o,i} \\
{\rm {\bf k}}\times {\rm {\bf E}} &=&\frac{\omega }{c}{\rm {\bf B}};
\end{eqnarray*}%
by elimination of ${\rm {\bf B}}$,

\begin{equation}
\Lambda _{ij}(\omega {\rm {\bf ,k}})E_{j}(\omega {\rm {\bf ,k}})=-\frac{4\pi
i}{\omega }j_{o,i}(\omega {\rm {\bf ,k}}),  \label{eq22}
\end{equation}%
where%
\begin{equation}
\Lambda _{ij}(\omega {\rm {\bf ,k}})=\frac{k^{2}c^{2}}{\omega ^{2}}(\frac{%
k_{i}k_{j}}{k^{2}}-\delta _{ij})+\varepsilon _{ij}(\omega {\rm {\bf ,k}}).
\label{eq23}
\end{equation}%
Multiplying Eq.(\ref{eq22}) by any modular vector $e_{{\rm {\bf k}}%
,i}^{\sigma \ast }$, with $e_{{\rm {\bf k}},i}^{\sigma }e_{{\rm {\bf k}}%
,i}^{\sigma \ast }=1,$ one obtains

\begin{equation}
\left( {k^2-\frac{\omega ^2}{c^2}\varepsilon _k^\sigma }\right) E_k^{T\sigma
}=\frac{4\pi i}{c^2}\omega [{\rm {\bf e}}_{{\rm {\bf k}}}^{\sigma *}\cdot
{\rm {\bf j}}^0(\omega ,{\rm {\bf k}})]\quad ,  \label{eq24}
\end{equation}
in which

\begin{equation}
\varepsilon _{k}^{\sigma }\equiv \varepsilon _{\omega ,{\rm {\bf k}}%
}^{\sigma }=\varepsilon _{ij}^{\sigma }(\omega ,{\rm {\bf k}})e_{{\rm {\bf k}%
},i}^{\sigma }e_{{\rm {\bf k}},j}^{\sigma \ast }+\frac{c^{2}}{\omega }({\rm
{\bf k}}\cdot {\rm {\bf e}}_{{\rm {\bf k}}}^{\sigma })({\rm {\bf k}}\cdot
{\rm {\bf e}}_{{\rm {\bf k}}}^{\sigma \ast });  \label{eq25}
\end{equation}%
and $\varepsilon _{k}^{\sigma }$is the \textquotedblleft dielectric
susceptibility\textquotedblright . By neglecting the terms in right hand (
the nonliear terms), one gets from Eq.(\ref{eq24}) the homogeneous wave
equation; and from here one finds the dispersion equation

\begin{equation}
\varepsilon _{k}^{\sigma }-\frac{k^{2}c^{2}}{\omega ^{2}}=0\quad .
\label{eq26}
\end{equation}%
Let $\omega =\omega ^{\sigma }({\rm {\bf k}})$ be its a solution ;
physically, the progressive wave in the opposite direction is also possible,
thus Eq.(\ref{eq26}) has another solution with negative frequency $\omega
=-\omega ^{\sigma }(-{\rm {\bf k}})$.

Consider the lowest order equation of motion for a test particle

\[
\quad \frac{d{\rm {\bf v}}}{dt}\approx {\rm {\bf E}}_g=-4\sqrt{G}{\rm {\bf E}%
},
\]
for plane waves $\propto e^{-i\omega t+i{\rm {\bf k}}\cdot {\rm {\bf r}}}$,
one has : $-i\omega {\rm {\bf v}}_{{\rm {\bf k}},\omega }=-4\sqrt{G}{\rm
{\bf E}}_{{\rm {\bf k}},\omega }$, thus

\[
{\rm {\bf j}}_{{\rm {\bf k}},\omega }=\sqrt{G}\rho {\rm {\bf v}}_{{\rm {\bf k%
}},\omega }=-\frac{\sqrt{G}\rho }{-i\omega }4\sqrt{G}{\rm {\bf E}}_{{\rm
{\bf k}},\omega }=\sigma {\rm {\bf E}}_{{\rm {\bf k}},\omega },
\]
i.e.

\[
j_{k,i}=\sigma _{ij}E_{k,j}{,\quad \sigma }_{ij}=\sigma \delta _{ij}=4\frac{%
G\rho }{i\omega }\delta _{ij}.
\]%
For transverse mode (${\rm {\bf k}}\cdot {\rm {\bf e}}_{{\rm {\bf k}}}^{t}=0)
$, on the other hand , one has from Eq.(\ref{eq20}) and Eq.(\ref{eq25})

\begin{equation}
\varepsilon _{k}^{t}=1+i\frac{\pi \sigma }{\omega }=1+\frac{4\pi G\rho }{%
\omega ^{2}};  \label{eq27}
\end{equation}%
with the \textquotedblleft gravitational conductivity\textquotedblright\ $%
\sigma $,

\[
\sigma =\sigma _{ij}(\omega ,{\rm {\bf k}})e_{{\rm {\bf k}},i}^{t}e_{{\rm
{\bf k}},j}^{t\ast }.
\]%
Hence from Eq.(\ref{eq26}) we obtain the dispersion relation

\begin{equation}
\begin{array}{l}
\omega ^{2}=k^{2}c^{2}-\omega _{0}^{2},{\quad \omega }_{0}^{2}\equiv 4\pi
G\rho ; \\
\omega \approx kc{\ \quad (\omega \gg \omega }_{0}{).}%
\end{array}
\label{eq28}
\end{equation}%
To calculate this emission radiated by the moving external currents, we form
the scalar product the first equation of Eq.(\ref{eq14}) with ${\rm {\bf E}}$%
, of the second one with ${\rm {\bf B}}$ and take the difference; then
integrating over a volume and using the Gauss theorem, as a result we obtain

\[
\frac{1}{4\pi }\int {[{\rm {\bf E}}\cdot \frac{\partial {\rm {\bf D}}}{%
\partial t}+{\rm {\bf B}}\cdot \frac{\partial {\rm {\bf B}}}{\partial t}]}d%
{\rm {\bf r}}+\frac{c}{4\pi }\int_{\sigma }{\left( {{\rm {\bf E}}\times {\rm
{\bf B}}}\right) d{\rm {\bf S}}}=-\int {{\rm {\bf j}}_{0}\cdot {\rm {\bf E}}d%
{\rm {\bf r}}};
\]%
taking into account that the ${\rm {\bf E}}$ and ${\rm {\bf B}}$ vanish at
infinity, thus

\begin{equation}
\frac{\partial W}{\partial t}\equiv \frac{1}{4\pi }[{\rm {\bf E}}\cdot \frac{%
\partial {\rm {\bf D}}}{\partial t}+{\rm {\bf B}}\cdot \frac{\partial {\rm
{\bf B}}}{\partial t}]=-{\rm {\bf j}}_{0}\cdot {\rm {\bf E}}.  \label{eq29}
\end{equation}%
i.e., the work done on the gravitoelectric fields per unit time, by the
external currents, makes the GEM field energy density ( $\left[ {j_{0}E}%
\right] \sim \left[ j_{0g}E_{g}/G\right] \sim \left[ {\rho
v^{2}}\right] /s)$ get increments. Hence the GEM wave energy
radiated is

\begin{equation}
U^{\sigma }=-\int {dt}\int {d{\rm {\bf r}}}{\rm {\bf E}}\left( {{\rm {\bf r}}%
,t}\right) \cdot {\rm {\bf j}}_{0}\left( {{\rm {\bf r}},t}\right) .
\label{eq30}
\end{equation}%
Using the Fourier transforms of ${\rm {\bf E}}$ and ${\rm {\bf j}}_{0}$ and
taking account of the condition of real field, we obtain the expression

\begin{equation}
U^{\sigma }=-\left( {2\pi }\right) ^{4}Re\int {d\omega d{\rm {\bf kj}}%
_{0}\left( {\omega ,{\rm {\bf k}}}\right) }\cdot {\rm {\bf E}}^{\ast }\left(
{\omega ,{\rm {\bf k}}}\right) .  \label{eq31}
\end{equation}%
In view of Eq.(\ref{eq24}),

\[
U^\sigma =2\left( {2\pi }\right) ^5Re\int {d\omega d{\rm {\bf k}}\frac{%
i\omega c^{-2}\left| {{\rm {\bf e}}^{\sigma *}({\rm {\bf k}})\cdot {\rm {\bf %
j}}_0\left( {\omega ,{\rm {\bf k}}}\right) }\right| ^2}{\left( {k^2-\frac{%
\omega ^2}{c^2}\varepsilon _k^\sigma }\right) ^{*}}};
\]
by use of the Landau rule and the Plemelj formula [15],
\[
Im\frac{1}{\left( {k^2 - \frac{\omega ^2}{c^2}\varepsilon _k^\sigma
} \right)^ * } = \mathop {Im}\limits_{\varepsilon \to 0}
\frac{1}{k^2 - \frac{\omega ^2}{c^2}\varepsilon _k^\sigma +
i\varepsilon } = - \pi \delta \left( {k^2 - \frac{\omega
^2}{c^2}\varepsilon _k^\sigma } \right)
\]
one has

\[
U^\sigma = \left( {2\pi } \right)^6\int {d{\rm {\bf k}}} \int_{ - \infty
}^\infty {d\omega \omega c^{ - 2}\left| {{\rm {\bf e}}^{\sigma \ast }({\rm
{\bf k}}) \cdot {\rm {\bf j}}(\omega ,{\rm {\bf k}})} \right|^2} \delta
\left( {k^2 - \frac{\omega ^2}{c^2}\varepsilon _k^\sigma } \right)
\]

\begin{equation}
=2\left( {2\pi }\right) ^6\int {d{\rm {\bf k}}\frac{\left| {{\rm {\bf e}}%
^{\sigma *}({\rm {\bf k}})\cdot {\rm {\bf j}}(\omega ^\sigma ({\rm {\bf k}}),%
{\rm {\bf k}})}\right| ^2}{\frac 1{\omega ^\sigma ({\rm {\bf k}})}\left| {%
\frac{\partial (\omega ^2\varepsilon _k^\sigma )}{\partial \omega }}\right|
|_{\omega =\omega ^\sigma ({\rm {\bf k}})}}},  \label{eq32}
\end{equation}
where

\begin{equation}
\varepsilon _{k}^{\sigma }\equiv \varepsilon _{\omega ,{\rm {\bf k}}%
}^{\sigma }=\varepsilon _{ij}^{\sigma }(\omega ,{\rm {\bf k}})e_{{\rm {\bf k}%
},i}^{\sigma }e_{{\rm {\bf k}},j}^{\sigma \ast };  \label{eq33}
\end{equation}%
the second term of Eq.(\ref{eq25}) has no contribution to the integration in
Eq.(\ref{eq32}); and in the last step we consider the solutions $\omega =\pm
\omega ^{\sigma }({\rm {\bf k}})$ of the dispersion equation, $k^{2}-\frac{%
\omega ^{2}}{c^{2}}\varepsilon _{k}^{\sigma }=0,$ have the same contribution
to the integration.

On the one hand , one may consider any instantaneous velocity as
uniform motion at short time intervals , and on the other hand the
space should be approximately flat for the slow-motion and weak
gravity, hence it is reasonable to study the emission radiated by a
neutron particle with mass $m$ in constant rectilinear motion
passing through the medium. Then the current density is

\begin{equation}
{\rm {\bf j}}_{0}({\rm {\bf r}},t)=q_{0}{\rm {\bf v}}(t)\delta \left( {{\rm
{\bf r}}-{\rm {\bf r}}(t)}\right) ,  \label{eq34}
\end{equation}%
where $q_{0}=\sqrt{G}m$, ${\rm {\bf r}}(t)$ is the position vector of the
motion particle. Considering that ${\rm {\bf r}}={\rm {\bf r}}_{0}+{\rm {\bf %
v}}t,\quad {\rm {\bf v}}=const.,$ the Fourier transform of Eq.(\ref{eq34}) is

\begin{equation}
{\rm {\bf j}}_{0}(\omega ,{\bf k})=\frac{q_{0}}{(2\pi )^{3}}{\bf v}\exp (i%
{\bf k\cdot r}_{0})\delta (\omega -{\bf k\cdot v}).  \label{eq35}
\end{equation}%
Because of the delta-function in Eq.(\ref{eq35}), for the emission by a
uniformly moving neutron particle has to be satisfied the Cerenkov
condition: $\omega ^{\sigma }={\rm {\bf k}}\cdot {\rm {\bf v}}$, or $%
v>\omega ^{\sigma }/k=c/n(\omega )$, which essentially corresponds to the
laws of energy and momentum conservation in the radiation processes. In view
of Eq.(\ref{eq27}), then $n(\omega )>1$; therefore, the Cerenkov emission by
the neutral particles in constant rectilinear motion is unavoidable.

Substituting the equation Eq.(\ref{eq35}) into Eq.(\ref{eq32}) the foregoing
results yield readily

\begin{equation}
U^\sigma =\int {\frac{d{\rm {\bf k}}}{(2\pi )^3}}\left[ {2(2\pi )^3\frac{%
\left| {{\rm {\bf e}}^{\sigma *}({\rm {\bf k}})\cdot {\rm {\bf j}}(\omega
^\sigma ,{\rm {\bf k}})}\right| ^2}{\frac 1{\omega ^\sigma ({\rm {\bf k}}%
)}\frac \partial {\partial \omega }\left[ {\omega ^2\varepsilon (\omega ,%
{\rm {\bf k}})}\right] _{\omega =\omega ^\sigma }}}\right] \quad .
\label{eq36}
\end{equation}
Using the following identity of delta-function
\[
\left[ {\delta \left( \omega \right)} \right]^2 = \mathop {\lim
}\limits_{\tau \to \infty } \frac{\tau }{2\pi }\delta \left( \omega
\right)
\]

\noindent and considering the definition of radiated power,
$I^\sigma = \mathop {\lim }\limits_{\tau \to \infty } \frac{U^\sigma
}{\tau }$£¬we obtain

\begin{equation}
I^{t}=\int {\frac{2\omega ^{t}({\rm {\bf k}}){\em q}_{0}^{2}\left\vert {{\rm
{\bf e}}^{t\ast }({\rm {\bf k}})\cdot {\rm {\bf v}}}\right\vert ^{2}}{\frac{d%
}{d\omega }\left[ {\omega ^{2}}{\em n}{^{2}(\omega )}\right] _{\omega
=\omega ^{t}}}}\delta (\omega ^{t}-{\rm {\bf k}}\cdot {\rm {\bf v}})\frac{d%
{\rm {\bf k}}}{2\pi }.  \label{eq37}
\end{equation}%
For transverse waves the polarization vector ${\rm {\bf e}}^{t}({\rm {\bf k}}%
)$ in Eq.(\ref{eq37}) has two independent components (${\rm {\bf e}}^{(1)},%
{\rm {\bf e}}^{(2)})$, then the total power radiated is obtain by adding the
two parts, i.e., proportional to $\left( {\left\vert {{\rm {\bf e}}^{(1)}(%
{\rm {\bf k}})\cdot {\rm {\bf v}}}\right\vert ^{2}+\left\vert {{\rm {\bf e}}%
^{(2)}({\rm {\bf k}})\cdot {\rm {\bf v}}}\right\vert ^{2}}\right) $; and
because ${\rm {\bf e}}^{(1)},{\rm {\bf e}}^{(2)}$ and ${\rm {\bf \hat{k}}}%
(\equiv {\rm {\bf k}}/k)$ constitute orthogonal frame in the complex space, $%
e_{i}^{\left( 1\right) }e_{j}^{\left( 1\right) }{}^{\ast }+e_{i}^{\left(
2\right) }e_{j}^{\left( 2\right) }{}^{\ast }=\delta _{ij}-k_{i}k_{j}/k^{2}$,
one has

\[
\left\vert {{\rm {\bf e}}^{(1)}({\rm {\bf k}})\cdot {\rm {\bf v}}}%
\right\vert ^{2}+\left\vert {{\rm {\bf e}}^{(2)}({\rm {\bf k}})\cdot {\rm
{\bf v}}}\right\vert ^{2}={\rm {\bf v}}^{2}-({\rm {\bf \hat{k}}}\cdot {\rm
{\bf v}})^{2}=v^{2}\sin ^{2}\theta ;
\]%
hence

\begin{equation}
I^{t}=\frac{Gm^{2}v}{c^{2}}\int_{v>\frac{c}{n(\omega )}}{d\omega \omega
\left( {1-\frac{c^{2}}{{\em v}^{2}n^{2}\left( \omega \right) }}\right) }.
\label{eq38}
\end{equation}%
It is well known that the gravitational system can emit energy in quadrupole
radiation; and the power radiated is proportional to the order of $c^{-5}$.
But one see from Eq.(\ref{eq38}) that the Cerenkov emission occurs in the
order of $c^{-2}$. Therefore, in general, the Cerenkov emission is more than
the quadrupole radiation. For example, consider a binary system, in which
the accreting compact object is a larger black hole with mass $%
M=10^{8}M_{\odot }$ and companion star with mass $m=M_{\odot }$ moves in
nearly circular orbit around the black hole at the radius $a_{0}=300r_{g}$ ($%
r_{g}=2GM/c^{2}$, the Schwarzschild radius ) . In such a case, from the
Kepler law, $(2\pi /p)^{2}=G(M+m)/a_{0}^{3}$, then one gets $p=1.56\times
10^{4}(h)$; at the same time, we may take$v\sim (2GM/a_{0})^{1/2}\sim
1.6\times 10^{9}(cm\cdot s^{-1})$. Furthermore, one can estimate the medium
density at $r=a_{0}$, $\rho =2.9\times 10^{-8}(a_{0}/3r_{g})^{3/2}=2.9\times
10^{-5}(g\cdot cm^{-3})$ [16]. As a result, due to the quadrupole radiation,
the energy loss for the system is [17]
\[
\frac{{\rm d}{\cal E}}{{\rm d}t}=3.0\times 10^{33}(\frac{\mu }{M_{\odot }}%
)^{2}(\frac{M}{M_{\odot }})^{4/3}(\frac{p}{1hour})^{-10/3}\approx 1.4\times
10^{30}(erg\cdot s^{-1}),
\]%
here $\mu =Mm/(M+m)$. And one gets from Eq.(\ref{eq38}) the energy loss in
Cerenkov radiation
\[
I^{\prime }=3\times 10^{33}\left( \frac{m}{M_{\odot }}\right) ^{2}\left(
\frac{v}{10^{9}cm\cdot s^{-1}}\right) \left( \frac{\omega }{10^{-7}}\right)
^{2}\approx 4.8\times 10^{30}(erg\cdot s^{-1});
\]%
we have taken $\omega \approx 10^{-7}(s^{-1})$ at the above. Then $%
n^{2}=\varepsilon ^{t}\simeq \frac{4\pi G\rho }{\omega ^{2}}=2.5\times
10^{3} $, thus the Cerenkov condition is satisfied: $v>c/n(\omega )$.

\bigskip

\noindent ACKNOWLEDGMENT

\noindent This work was partly supported by the National Natural Science
Foundation of China and the Natural Science Foundation of Jiangxi Province

\bigskip

\begin{flushleft}{\bf REFERENCES}\end{flushleft}
\bigskip [1]Shapiro S.(1996). {\em Phys.Rev.Lett}. {\bf 77} , 4487. \newline
\noindent[2]Bonilla M.A.G.,and SenovillaJ.M.M.(1997).{\em Phys.Rev.Lett}.
{\bf 78},783. \newline
\noindent[3] Senovilla J.M.M.(2000).{\em Mod.Phys.Lett}. {\em A}{\bf 151},59.%
\newline
\noindent[4] Ruggiero M.L., and Tartaglia A.(2002). {\em Nuovo Cimento B}
{\bf 117},743.\newline
\noindent[5] Iorio L., and Lucchesi D.M.(2003). {\em Class. Quantum Grav}.%
{\bf 20}, 2477.\newline
\noindent[6] Melrose D.B. (1986).{\em Instabilities in space and laboratory
plasmas},\newline
\indent ~~~~Cambrige University Press,Cambrige.\newline
\noindent[7] Weinberg S. (1972). {\em Gravitation and Cosmology}, John Wiley
{\&} Sons, Inc,New York.\newline
\noindent[8] Braginsky V.B., Caves C.M., and Thorne Kip S. (2003). {\em %
Phys.Rev. D}{\bf 15}, 2045.\newline
\noindent[9] Braginsky V.B. , Polnarev A.G. , and Thorne Kip S.(1984). {\em %
Phys.Rev.Lett}.{\bf 53}, 863.\newline
\noindent[10] Ciubotariu C.(1991). {\em Phys.Lett.A} {\bf 158}, 27.\newline
\noindent[11] Agop M., Buzea C.Gh., and Nica P.(2000). {\em Physica C }{\bf %
339}, 120.\newline
\noindent[12] Oprea I., and Agop M.(1998). {\em Studia Geoph.Et Geod}. {\bf %
42}, 431.\newline
\noindent[13] Li X.Q.(1990). {\em Astron.Astrophys}.{\bf 227}, 317.\newline
\noindent[14] Li X.Q., and Ma Y.H.(1993). {\em Astron.Astrophys}.{\bf 270},
534.\newline
\noindent[15] Lifshitz E.M. , and Pitaevskii L.P.(1981). {\em Physical
Kinetics}, Pergamon, Oxford.\newline
\noindent[16] Takahara F.(1979). {\em Prog.Theor.Phys}. {\bf 62}, 629.%
\newline
\noindent[17] Peters P. C. and Mathews J.(1963). {\em Phys.Rev}.{\bf 131},
435.\newline

\end{document}